\documentclass[aps,prl,twocolumn,groupedaddress,showpacs,showkeys]{revtex4}

\usepackage{graphicx}% Include figure files
\usepackage{dcolumn}% Align table columns on decimal point
\usepackage{bm}% bold math
\usepackage{amssymb}
\usepackage{amsmath}

\usepackage{color}

\newcommand{\todo}[1]{\textcolor{blue}{#1}}
\renewcommand{\todo}[1]{#1}
\begin{document}

% Use the 'preprint numbers' class option to override journal defaults
%\preprint{}

\title{FIRST EXPERIMENTAL OBSERVATION OF GENERALIZED SYNCHRONIZATION
PHENOMENA IN MICROWAVE OSCILLATORS}


\author{Boris~S.~Dmitriev}
%\email{DmitrievBS@info.sgu.ru}
\author{Alexander~E.~Hramov}
%\email{aeh@nonlin.sgu.ru}
\author{Alexey~A.~Koronovskii}
%\email{alkor@nonlin.sgu.ru}
\author{Andrey~V.~Starodubov}
%\email{StarodubovAV@nonlin.sgu.ru}
\author{Dmitriy~I.~Trubetskov}
%\email{TrubetskovDI@nonlin.sgu.ru}
\author{Yurii~D.~Zharkov}
%\email{ZharkovYD@info.sgu.ru}
\affiliation{Faculty of Nonlinear
Processes, Saratov State University, Astrakhanskaya, 83, Saratov,
410012, Russia}

%\thanks{}


\date{\today}

\begin{abstract}
In this Letter we report for the first time on the experimental
observation of the generalized synchronization regime in the
microwave electronic systems, namely, in the multicavity klystron
generators. A new approach devoted to the generalized
synchronization detection has been developed. The experimental
observations are in the excellent agreement with the results of
numerical simulation. The observed phenomena gives a strong
potential for new applications requiring microwave chaotic signals.
\end{abstract}


\maketitle


Chaotic synchronization is one of the fundamental phenomena, widely
studied recently, having both theoretical and applied
significance~\cite{Boccaletti:2002_ChaosSynchro}. One of the
interesting and intricate types of the synchronous behavior of
unidirectionally coupled chaotic oscillators is the generalized
synchronization (GS)~\cite{Rulkov:1996_AuxiliarySystem,
Aeh:2005_GS:ModifiedSystem}, which means the presence a functional
relation between the dynamics of the drive and response chaotic
systems, though this relation may be very complicated and its
explicit form cannot be found in most cases. Remarkably, that
practically all studies of the generalized synchronization
phenomenon deal with the low-dimensional model systems or the low
frequency oscillators. Even if generalized synchronization in lasers
is studied, the oscillations of the total intensity of the laser
output are usually considered whose frequency is in the megahertz
range, with the oscillator dynamics being described by the system of
the ordinary differential equations~\cite{Uchida:2003_GSLaserPRL}.
The more complicated objects with the infinite dimensional phase
space (such as spatially extended systems  (see, e.g.,
\cite{Hramov:2005_GLEsPRE}) or oscillators with the delayed
feedback) are considered from the point of view of generalized
synchronization rarely, and, practically always, numerically.

In this Letter we report for the first time on the experimental
revelation of generalized synchronization in the microwave systems
with the infinite dimensional phase space, namely, in the
multicavity klystron oscillators with the delayed feedback. Along
with the theoretical interest this study is also important from the
point of view of the practical purposes of communication, where the
microwave range signals are used very widely.

To detect the onset of generalized synchronization in the experiment
a new approach being applicable to the microwave systems has been
developed. Actually, there are various techniques for detecting the
presence of GS between chaotic oscillators, such as the method of
nearest neighbors~\cite{Rulkov:1995_GeneralSynchro} or the auxiliary
system approach~\cite{Rulkov:1996_AuxiliarySystem}. It is also
possible to calculate the conditional Lyapunov
exponents~\cite{Pyragas:1996_WeakAndStrongSynchro} to reveal the
generalized synchronization regime. Unfortunately, these approaches
are often difficult (or impossible) to implement in the experimental
measurements (specifically, in the microwave range), due to the
presence of noise and lack of the precisions. Therefore, in this
Letter we propose the radically different approach which may deal
with the chaotic oscillators of the microwave range to detect
experimentally the onset of the generalized synchronization regime.
Moreover, as we show below, the developed technique may be also used
successfully to detect the onset of the GS regime for a very wide
range of dynamical systems outside the microwave electronics.

\begin{figure}[b]
%\vspace*{4cm}
\centerline{\includegraphics*[scale=0.45]{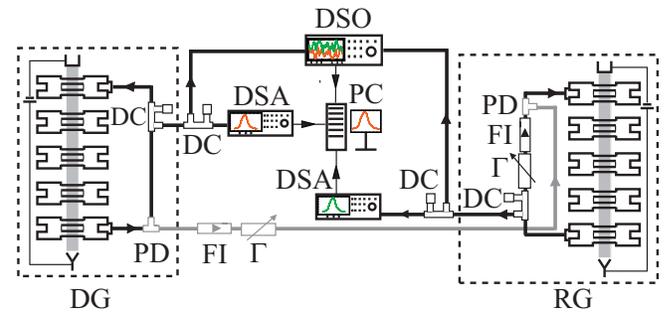}} \caption{(Color
online) The experimental setup for the GS observation in two
unidirectionally coupled microwave klystron generators with the
delayed feedback. The dashed rectangles correspond to the drive (DG)
and response (RG) generators. FI --- ferrite isolator; PD --- power
divider; DC --- directional coupler; $\Gamma$ --- waveguide
attenuator; DSO --- digital storage oscilloscope Agilent
Technologies DSO81004B; DSA --- digital spectrum analyzer Agilent
Technologies E4402B; PC --- computer. The control parameters of the
drive (labeled by ``\textit{d}'') and response (labeled by
``\textit{r}'') oscillators are $V_{0}^{d}=2.1$~kV, $I^{d}=50$~mA,
$V_{0}^{r}=1.9$~kV, $I^{r}=52$~mA} \label{fgr:ExperimetalSetup}
\end{figure}

\begin{figure}[tb]
%\vspace*{4cm}
\centerline{\includegraphics*[scale=0.4]{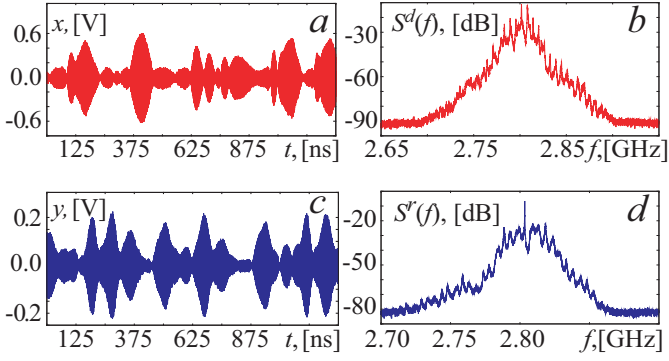}} \caption{(Color
online) The experimentally obtained time series and power spectra.
Figures \textit{a}, \textit{b} correspond to the drive oscillator
and \textit{c}, \textit{d} --- to the response one. Both oscillators
are in the autonomous regime}\label{fgr:ExperimetalSpectra}
\end{figure}

Our experimental setup is shown in Fig.~\ref{fgr:ExperimetalSetup}.
We use the S-band five-cavity floating-drift klystron amplifier with
the delayed feedback~\cite{Shigaev:2005_Klystron} as a drive
generator (DG). We can tune the behavior of the generator by
changing the acceleration voltage $V_{0}$ and the electron beam
current $I_0$. The output signal of the drive system with the help
of the coaxial line (shown in gray) is transmitted to the input of
the response generator (RG), which is analogous to the drive one. To
prevent the backward influence of the response klystron oscillator
to the drive system we include the ferrite isolator (FI) into the
coupling line. The coupling strength between interacting generators
may be regulated within wide limits with the help of the attenuators
($\Gamma$). The output microwave signals of both drive [$x(t)$] and
response [$y(t)$] generators are measured by the digital storage
oscilloscope (DSO) and the digital spectrum analyzer (DSA) and
stored in a computer (PC). The typical fragments of time series and
power spectra obtained experimentally for the considered klystron
generators are shown in Fig.~\ref{fgr:ExperimetalSpectra}.

To detect the onset of the GS regime we propose to analyze the power
spectra of chaotic oscillations being one of the important
characteristics of the system dynamics. Additionally, they are
easily obtained in the experiments. The main idea of the proposed
approach is the following. Since the chaotic oscillator dynamics
changed sufficiently when GS
arises~\cite{Aeh:2005_GS:ModifiedSystem}, one can expect that this
transformation is also manifested in the power spectra
modifications. To detect the qualitative changes in the power
spectrum of the response system taking place with the increase of
the coupling strength ${\varepsilon=10^{-\Gamma/20}}$ (where
$\Gamma$ is the attenuation factor of the coupling microwave line)
we propose the following characteristic
\begin{equation}
\begin{array}{l}
\displaystyle{\sigma(\varepsilon)=\frac{1}{P^{d}}\int\limits_{0}^{\infty}
\left(\frac{\partial S^{r}(f,\varepsilon)}{\partial \varepsilon}\right)^{2}df},\\
\end{array}
\label{eq:IntegralMeasure}
\end{equation}
where $P^{d}=\int_{0}^{\infty}S^{d}(f)\,df$ is the total power of
the spectrum $S^{d}(f)$ of the drive generator oscillations and
$S^{r}(f,\varepsilon)$ --- the power spectrum of the response
oscillator observed for the coupling strength $\varepsilon$. The
derivative $\partial S^{r}(f,\varepsilon)/\partial\varepsilon$
specifies the changes of the power spectra of the response system
when the coupling strength $\varepsilon$ is changed. Unless the
dynamics of the response system is transformed cardinally the value
$\sigma(\varepsilon)$ is not supposed to be changed greatly. Since
arising the GS regime means the major restructuring of the response
system dynamics~\cite{Aeh:2005_GS:ModifiedSystem}, one can expect
that the noticeable variations in the evolution of the
characteristic $\sigma(\varepsilon)$ corresponding to the onset of
the GS regime may be observed.
Since the power spectra obtained in the experiment are represented
by the discrete sets of data, the integral
in~(\ref{eq:IntegralMeasure}) should be replaced by the sum and,
therefore, Eq.~(\ref{eq:IntegralMeasure}) takes the form
\begin{equation}\nonumber
\begin{array}{l}
\displaystyle{\sigma(\varepsilon)=\frac{1}
{\Delta\varepsilon^{2}P^{d}}\sum^{N}_{i=0}\left(\langle S^{r}(f_{i},
\varepsilon)\rangle-\langle S^{r}(f_{i},\varepsilon-\Delta\varepsilon)\rangle\right)^{2}},\\
\end{array}
\label{eq:numerical_mera GS}
\end{equation}
where $P^{d}=\sum_{i=1}^{N}\langle (S^{d}(f_i))^{2}\rangle$,
${N=2^{12}}$ is the number of the spectral components in the
discrete representation of the power spectrum, $\langle\cdot\rangle$
denotes the ensemble average. In the experiment we have used
averaging over 512 measurements.

\begin{figure}[tb]
\centerline{\includegraphics*[scale=0.4]{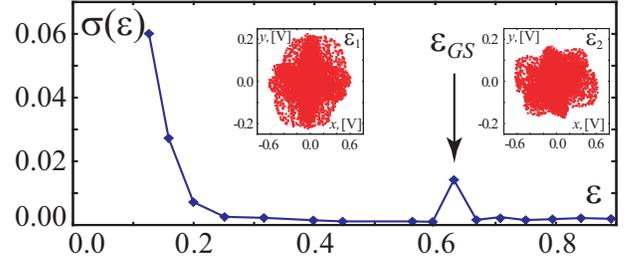}} \caption{(Color
online) The experimental characteristic $\sigma(\varepsilon)$ for
two unidirectionally coupled klystron generators. The correlation
plots $y$ vs. $x$ are shown in the frames for $\varepsilon_1=0.10$
and $\varepsilon_2=0.89$, respectively.}\label{fgr:characteristics}
\end{figure}

The characteristics $\sigma(\varepsilon)$ obtained from the
experiment is shown in Fig.~\ref{fgr:characteristics}. One can see
that after the monotonous decrease the curve $\sigma$ vs.
$\varepsilon$ shows the pronounced peak
($\varepsilon_{GS}\approx0.63$) being the manifestation of the great
transformation of the response oscillator dynamics. This
transformation of the response system behavior is expected to be
associated with the onset of the GS regime. Nevertheless, we have to
be convinced that the observed peak is really the evidence of the GS
regime onset. Note, there is no complete synchronization between
oscillators (see correlation plots $y$ vs. $x$ in frames in
Fig.~\ref{fgr:characteristics}).

\begin{figure}[tb]
\centerline{\includegraphics*[scale=0.4]{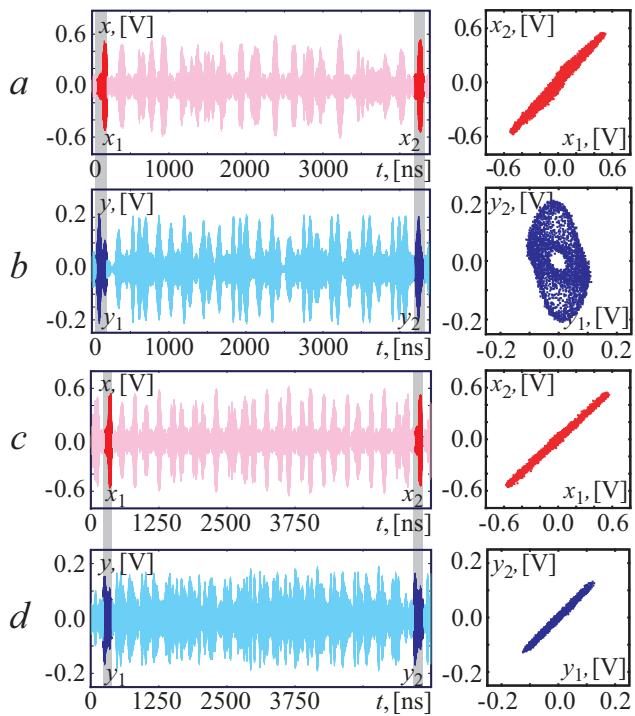}} \caption{(Color
online) The selected fragments of time series and corresponding to
them correlation plots. Figures \textit{a} (drive) and \textit{b}
(response) correspond to the coupling strength $\varepsilon_1=0.10$,
whereas \textit{c} (drive) and \textit{d} (response) --- to the
coupling strength $\varepsilon_2=0.89$}\label{fgr:CorrelationPlot}
\end{figure}

To verify the assumption made above, we consider the experimental
time series of the drive and response oscillators for two different
values of the coupling strength $\varepsilon_1=0.10$ and
$\varepsilon_2=0.89$, below and above the critical value
$\varepsilon_{GS}$ supposed to be the boundary of GS. Having
selected on different time intervals
${\mathfrak{T}_1:t\in[t_1,t_1+T)}$ and
${\mathfrak{T}_2:t\in[t_2,t_2+T)}$ (where $T=130$~ns is the time
delay of the generator feedback) two nearly identical pieces
$x_1(\xi)$, $x_2(\xi)$ of the drive oscillator time series and
corresponding to them two segments $y_1(\xi)$, $y_2(\xi)$ of the
response system time series, we draw the correlation plots $x_2$ vs.
$x_1$ and $y_2$ vs. $y_1$, respectively
(Fig.~\ref{fgr:CorrelationPlot}). To detect such similar segments of
the drive system time series we consider the distances
$d=\int_{0}^T(x_1(\xi)-x_2(\xi))^2d\xi$ between two arbitrary pieces
$x_1$, $x_2$ with length $T$ and find the minimal value of $d$.
Obviously, if the drive system behavior is practically identical on
time intervals $\mathfrak{T}_1$ and $\mathfrak{T}_2$ (see
Fig.~\ref{fgr:CorrelationPlot},\,\textit{a,c}), the response system
shows also nearly identical dynamics when the GS regime takes place
(Fig.~\ref{fgr:CorrelationPlot},\,\textit{d}), whereas in the case
of absence of GS the behavior of the response oscillator on
$\mathfrak{T}_1$ and $\mathfrak{T}_2$ is different
(Fig.~\ref{fgr:CorrelationPlot},\,\textit{b}).

In parallel with the experimental study we have also carried out a
numerical simulations of a simple model of two unidirectionally
coupled klystron generators with the delayed feedback. With the
dynamics of the autonomous klystron generator being in accordance
with the numerical simulations of this
model~\cite{Shigaev:2005_Klystron}, the qualitative agreement
between experimental and numerical results concerning the behavior
of two unidirectionally coupled klystron generators with delay is
expected to take place, too. The mathematical form of two coupled
generator model is the following
\begin{equation}
\begin{array}{l}
\displaystyle{\dot{F}^{d}_{1}(\tau)+\gamma^{d}{F}^{d}_{1}(\tau)=\gamma^{d}{F}^{d}_{2}(\tau-1)},\\
\displaystyle{\dot{F}^{d}_{2}(\tau)+\gamma^{d}{F}^{d}_{2}(\tau)=-2i\alpha^{d}{e}^{-i\psi}J_{1}(|F^{d}_{1}(\tau)|)\frac{F^{d}_{1}(\tau)}{|F^{d}_{1}(\tau)|}},\\
\displaystyle{\dot{F}^{r}_{1}(\tau)+\gamma^{r}{F}^{r}_{1}(\tau)=\gamma^{r}((1-\varepsilon){F}^{r}_{2}(\tau-1)+\varepsilon{F}^{d}_{2}(\tau))},\\
\displaystyle{\dot{F}^{r}_{2}(\tau)+\gamma^{r}{F}^{r}_{2}(\tau)=-2i\alpha^{r}{e}^{-i\psi}J_{1}(|F^{r}_{1}(\tau)|)\frac{F^{r}_{1}(\tau)}{|F^{r}_{1}(\tau)|}},\\
\end{array}
\label{eq:KlystronModel}
\end{equation}
where indexes ``$d$'' and ``$r$'' relate to the drive and response
oscillators; $F_1$ and $F_2$ are the slowly varying dimensionless
amplitudes of oscillations in the input and output cavities,
respectively; $\tau$ is the normalized time, $\psi$ is the phase
shift during propagation along the feedback circuit, $J_1$ is a
Bessel function of the first kind, $\gamma$ is the loss parameter,
$\alpha$ is a beam current, $\varepsilon$ is a coupling strength.
The control parameter values are $\alpha^{r}=10.9$,
$\alpha^{d}=10.85$, $\psi=-0.5\pi$, $\gamma^{r}=1.0$,
$\gamma^{d}=1.0$. For such a choice of the control parameter values,
both the drive and response oscillators generate the chaotic
signals.

\begin{figure}[tb]
%\centerline{\includegraphics*[scale=0.3]{fig5.eps}}
\centerline{\includegraphics*[scale=0.4]{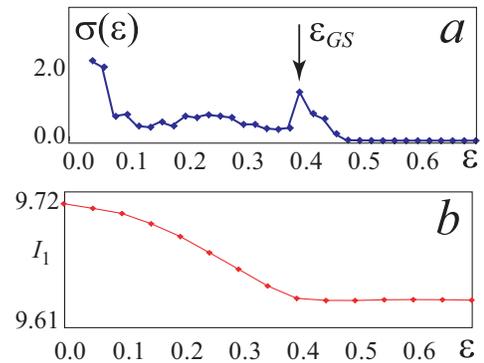}} \caption{(Color
online) (\textit{a}) The characteristic $\sigma(\varepsilon)$ for
two unidirectionally coupled klystron
generators~(\ref{eq:KlystronModel}) and (\textit{b}) the dependence
of the value of the first harmonic of the bunched electron beam
current $|I_{1}|$ in the response klystron oscillator on the
coupling strength $\varepsilon$. The arrows show the GS onset in
coupled microwave oscillators}\label{fgr:FH}
\end{figure}

\todo{The character of the evolution of $\sigma$ quantity
(Fig.~\ref{fgr:FH},\,\textit{a}) is quite similar both for the
experiment and numerical model.} The peak being analogous to the one
of the experimental curve happens to be also observed in the same
range of the coupling strength value for the plot
$\sigma(\varepsilon)$ obtained by means of numerical simulation of
the model system~(\ref{eq:KlystronModel}) (see
Fig.~\ref{fgr:FH},\,\textit{a}). Fortunately, the traditional
methods of the GS detection mentioned above may be used for the
numerical simulation data (contrary to the experimental ones). Based
on the auxiliary system approach~\cite{Rulkov:1996_AuxiliarySystem}
we have found that the peak on the curve $\sigma(\varepsilon)$ is
certainly the manifestation of the GS onset. Indeed, there is the
exact coincidence of the values of coupling strength $\varepsilon$
corresponding to the extremum of the peak of the curve
$\sigma(\varepsilon)$ and the onset of GS shown in
Fig.~\ref{fgr:FH},\,\textit{a} by an arrow. So, the peak on the
curve $\sigma(\varepsilon)$ really corresponds to the boundary of
the GS regime.

Let us discuss briefly the mechanism of arising the GS regime in a
system of unidirectionally coupled klystron oscillators. It is
well-known~\cite{Pyragas:1996_WeakAndStrongSynchro,Harmov:2005_GSOnset_EPL}
that the proper chaotic dynamics of the response system should be
suppressed for the GS regime to take place. The same reason results
in arising the GS regime in coupled klystron oscillators. Indeed,
the dynamics of the klystron generator is determined by the value of
the first harmonic of the bunched electron beam current. In the
context of the used model~(\ref{eq:KlystronModel}) this
dimensionless first harmonic of current may be written as
$I_{1}=2\alpha^r J_{1}\left(|F_{1}(\tau)|/2\right)$.

%\begin{equation}
%\begin{array}{l}
%\displaystyle{I_{1}=2 i I_{0}J_{1}\left(\frac{|F_{1}(\tau)|}{2}\right)e^{-i(\theta_{0}-\mathrm{arg}(F_{1}(\tau)))}}\\
%\end{array}
%\label{eq:first harmonic}
%\end{equation}
%where $I_{0}$ is the beam current, $\theta_{0}$ is the unperturbed
%electron transit angle in the drift space,
%$\mathrm{arg}(F_{1}(\tau))$ is the phase of oscillations in the
%first resonator.

The dependence of the value of the first harmonic of current on the
coupling strength $\varepsilon$ obtained by means of the numerical
simulations is shown in Fig.~\ref{fgr:FH},\,\textit{b}. One can see
that the curve $I_{1}(\varepsilon)$ decreases from the value
$I_{1}\approx9.72$ to $I_{1}\approx9.61$ when the coupling strength
$\varepsilon$ grows, with the sharp decrease corresponding to the GS
onset. The generalized synchronization arises when the first
harmonic arrives the value $I_{1}\approx9.61$ (see
Fig.~\ref{fgr:FH}).
%As we have
%found, only periodic oscillations may be observed for the autonomous
%klystron oscillator~(\ref{eq:KlystronModel}) when
%$|I_{1}/I_{0}|\approx0.73$.
Therefore, near the critical value of
the coupling strength $\varepsilon_c\approx0.4$ the influence of the
drive klystron oscillator results in the decrease of the first
harmonic of the current and, correspondingly, in the suppression of
the proper chaotic dynamics of the response system, with the GS
regime being revealed.

\begin{figure}[tb]
\centerline{\includegraphics*[scale=0.4]{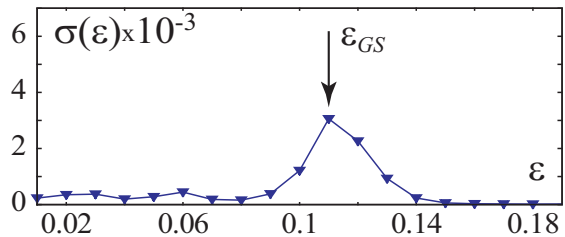}} \caption{The
characteristic $\sigma(\varepsilon)$ obtained from the numerical
simulations of unidirectionally coupled R\"ossler oscillators with
the diffusion type of coupling. The control parameters are $a=0.15$,
$p=0.2$, $c=10.0$, $w_{r}=0.95$, $w_{d}=0.99$. }\label{fgr:Roessler}
\end{figure}

There is one more important problem to be considered, namely,
whether the proposed GS onset detection technique associated with
the characteristic~(\ref{eq:IntegralMeasure}) is applicable for a
wide range of the different systems. To examine this problem we have
studied the dependence of the measure $\sigma(\varepsilon)$ on the
coupling strength $\varepsilon$ for two unidirectionally coupled
R\"ossler oscillators
\begin{equation}
\begin{array}{ll}
\displaystyle{\dot{x}_{d}=-w_{d}y_{d}-z_{d}}, & \displaystyle{\dot{x}_{r}=-w_{r}y_{r}-z_{r}+\varepsilon(x_{d}-x_{r})},\\
\displaystyle{\dot{y}_{d}=w_{d}x_{d}+ay_{d}}, & \displaystyle{\dot{y}_{r}=w_{r}x_{r}+ay_{r}},\\
\displaystyle{\dot{z}_{d}=p+z_{d}(x_{d}-c)}, & \displaystyle{\dot{z}_{r}=p+z_{r}(x_{r}-c)}.\\
\end{array}
\label{eq:Roesslers}
\end{equation}
In~(\ref{eq:Roesslers}) the indexes $``d"$ and $``r"$ relate to the
drive and response oscillators, $\varepsilon$ is a coupling
strength. The numerically obtained curve $\sigma(\varepsilon)$ is
shown in Fig.~\ref{fgr:Roessler}. As well as in the case of two
unidirectionally coupled klystron generators there is the pronounced
peak in the curves $\sigma(\varepsilon)$ coinciding with the point
where the GS regime arises (the onset of GS obtained by means of the
auxiliary system approach is shown by an arrow in
Fig.~\ref{fgr:Roessler}). The very same results have been obtained
for the unidirectionally coupled R\"ossler and Lorenz oscillators
and for the unidirectionally coupled complex Ginzburg-Landau
equations (The phenomenon of GS in the coupled Ginzburg-Landau
equations was described in~\cite{Hramov:2005_GLEsPRE}). \todo{The
obtained results provide a basis for expectation that the developed
technique can be also applied to bidirectional coupled systems, but
this problem deserves a further careful consideration going far
beyond the subject of the present Letter.}

In conclusion, we have reported on the first experimental
observation of the generalized synchronization phenomena in the
microwave oscillators with the delayed feedback. We have developed
the new techniques of GS determination that can be applied to the
experimental study. The proposed techniques have been tested both on
experimentally and numerically obtained data. The mechanism of GS
arising has been revealed. The experimental observations are in the
excellent agreement with the results of numerical simulation. The
observed phenomena is supposed to give a strong potential for new
applications requiring microwave chaotic signals.


\begin{thebibliography}{9}
\expandafter\ifx\csname natexlab\endcsname\relax\def\natexlab#1{#1}\fi
\expandafter\ifx\csname bibnamefont\endcsname\relax
  \def\bibnamefont#1{#1}\fi
\expandafter\ifx\csname bibfnamefont\endcsname\relax
  \def\bibfnamefont#1{#1}\fi
\expandafter\ifx\csname citenamefont\endcsname\relax
  \def\citenamefont#1{#1}\fi
\expandafter\ifx\csname url\endcsname\relax
  \def\url#1{\texttt{#1}}\fi
\expandafter\ifx\csname urlprefix\endcsname\relax\def\urlprefix{URL }\fi
\providecommand{\bibinfo}[2]{#2}
\providecommand{\eprint}[2][]{\url{#2}}

\bibitem[{\citenamefont{Boccaletti et~al.}(2002)\citenamefont{Boccaletti,
  Kurths, Osipov, Valladares, and Zhou}}]{Boccaletti:2002_ChaosSynchro}
\bibinfo{author}{\bibfnamefont{S.}~\bibnamefont{Boccaletti}},
  \bibinfo{author}{\bibfnamefont{J.}~\bibnamefont{Kurths}},
  \bibinfo{author}{\bibfnamefont{G.~V.} \bibnamefont{Osipov}},
  \bibinfo{author}{\bibfnamefont{D.~L.} \bibnamefont{Valladares}},
  \bibnamefont{and} \bibinfo{author}{\bibfnamefont{C.~T.} \bibnamefont{Zhou}},
  \bibinfo{journal}{Physics Reports} \textbf{\bibinfo{volume}{366}},
  \bibinfo{pages}{1} (\bibinfo{year}{2002}).

\bibitem[{\citenamefont{Abarbanel et~al.}(1996)\citenamefont{Abarbanel, Rulkov,
  and Sushchik}}]{Rulkov:1996_AuxiliarySystem}
\bibinfo{author}{\bibfnamefont{H.~D.} \bibnamefont{Abarbanel}},
  \bibinfo{author}{\bibfnamefont{N.~F.} \bibnamefont{Rulkov}},
  \bibnamefont{and} \bibinfo{author}{\bibfnamefont{M.~M.}
  \bibnamefont{Sushchik}}, \bibinfo{journal}{Phys. Rev. E}
  \textbf{\bibinfo{volume}{53}}, \bibinfo{pages}{4528} (\bibinfo{year}{1996}).

\bibitem[{\citenamefont{Hramov and
  Koronovskii}(2005)}]{Aeh:2005_GS:ModifiedSystem}
\bibinfo{author}{\bibfnamefont{A.~E.} \bibnamefont{Hramov}} \bibnamefont{and}
  \bibinfo{author}{\bibfnamefont{A.~A.} \bibnamefont{Koronovskii}},
  \bibinfo{journal}{Phys. Rev. E} \textbf{\bibinfo{volume}{71}},
  \bibinfo{pages}{067201} (\bibinfo{year}{2005}).

\bibitem[{\citenamefont{Uchida et~al.}(2003)\citenamefont{Uchida, McAllister,
  Meucci, and Roy}}]{Uchida:2003_GSLaserPRL}
\bibinfo{author}{\bibfnamefont{A.}~\bibnamefont{Uchida}},
  \bibinfo{author}{\bibfnamefont{R.}~\bibnamefont{McAllister}},
  \bibinfo{author}{\bibfnamefont{R.}~\bibnamefont{Meucci}}, \bibnamefont{and}
  \bibinfo{author}{\bibfnamefont{R.}~\bibnamefont{Roy}},
  \bibinfo{journal}{Phys. Rev. Lett.} \textbf{\bibinfo{volume}{91}},
  \bibinfo{pages}{174101} (\bibinfo{year}{2003}).

\bibitem[{\citenamefont{Hramov et~al.}(2005{\natexlab{a}})\citenamefont{Hramov,
  Koronovskii, and Popov}}]{Hramov:2005_GLEsPRE}
\bibinfo{author}{\bibfnamefont{A.~E.} \bibnamefont{Hramov}},
  \bibinfo{author}{\bibfnamefont{A.~A.} \bibnamefont{Koronovskii}},
  \bibnamefont{and} \bibinfo{author}{\bibfnamefont{P.~V.} \bibnamefont{Popov}},
  \bibinfo{journal}{Phys. Rev. E} \textbf{\bibinfo{volume}{72}},
  \bibinfo{pages}{037201} (\bibinfo{year}{2005}{\natexlab{a}}).

\bibitem[{\citenamefont{Rulkov et~al.}(1995)\citenamefont{Rulkov, Sushchik,
  Tsimring, and Abarbanel}}]{Rulkov:1995_GeneralSynchro}
\bibinfo{author}{\bibfnamefont{N.~F.} \bibnamefont{Rulkov}},
  \bibinfo{author}{\bibfnamefont{M.~M.} \bibnamefont{Sushchik}},
  \bibinfo{author}{\bibfnamefont{L.~S.} \bibnamefont{Tsimring}},
  \bibnamefont{and} \bibinfo{author}{\bibfnamefont{H.~D.}
  \bibnamefont{Abarbanel}}, \bibinfo{journal}{Phys. Rev. E}
  \textbf{\bibinfo{volume}{51}}, \bibinfo{pages}{980} (\bibinfo{year}{1995}).

\bibitem[{\citenamefont{Pyragas}(1996)}]{Pyragas:1996_WeakAndStrongSynchro}
\bibinfo{author}{\bibfnamefont{K.}~\bibnamefont{Pyragas}},
  \bibinfo{journal}{Phys. Rev. E} \textbf{\bibinfo{volume}{54}},
  \bibinfo{pages}{R4508} (\bibinfo{year}{1996}).

\bibitem[{\citenamefont{Shigaev et~al.}(2005)\citenamefont{Shigaev, Dmitriev,
  Zharkov, and Ryskin}}]{Shigaev:2005_Klystron}
\bibinfo{author}{\bibfnamefont{A.~M.} \bibnamefont{Shigaev}},
  \bibinfo{author}{\bibfnamefont{B.~S.} \bibnamefont{Dmitriev}},
  \bibinfo{author}{\bibfnamefont{Y.}~\bibnamefont{Zharkov}}, \bibnamefont{and}
  \bibinfo{author}{\bibfnamefont{N.~M.} \bibnamefont{Ryskin}},
  \bibinfo{journal}{IEEE Transactions on Electron Devices}
  \textbf{\bibinfo{volume}{52}}, \bibinfo{pages}{790} (\bibinfo{year}{2005}).

\bibitem[{\citenamefont{Hramov et~al.}(2005{\natexlab{b}})\citenamefont{Hramov,
  Koronovskii, and Moskalenko}}]{Harmov:2005_GSOnset_EPL}
\bibinfo{author}{\bibfnamefont{A.~E.} \bibnamefont{Hramov}},
  \bibinfo{author}{\bibfnamefont{A.~A.} \bibnamefont{Koronovskii}},
  \bibnamefont{and} \bibinfo{author}{\bibfnamefont{O.~I.}
  \bibnamefont{Moskalenko}}, \bibinfo{journal}{Europhysics Letters}
  \textbf{\bibinfo{volume}{72}}, \bibinfo{pages}{901}
  (\bibinfo{year}{2005}{\natexlab{b}}).

\end{thebibliography}
\end{document}